# Optimal Communication Channels in a Disordered World with Tamed Randomness

Philipp del Hougne[1]*, Mathias Fink[1], Geoffroy Lerosey[2]*

[1]Institut Langevin, CNRS UMR 7587, ESPCI Paris, PSL Research University, 1 rue Jussieu, 75005 Paris, France.

[2]Greenerwave, ESPCI Paris Incubator PC'up, 6 rue Jean Calvin, 75005 Paris, France.

*Correspondence to philipp.delhougne@gmail.com or geoffroy.lerosey@greenerwave.com.

**Multi-channel wireless systems have become a standard solution to address our information society's ever-increasing demand for more information transfer. The ultimate bound on the capacity such systems can achieve is the limited channel diversity in a given propagation medium, and countless proposals to reduce channel cross-talk by engineering software or hardware details of the signals and antenna arrays have been proposed. Here we demonstrate the physical shaping of the propagation medium itself to achieve optimal channel diversity. Using a reconfigurable metasurface placed inside a random environment, we tune the disorder and impose perfect orthogonality of wireless channels. We present experiments in the microwave domain in which we impose equal weights of the channel matrix eigenvalues for up to 4×4 systems, and almost equal weights in larger systems. We further describe one example of wireless image transmission in an office room in which we augmented the 3×3 system's number of effectively independent channels from 2 to the optimum of 3.**

Shannon defined a communication channel's maximum rate of information transfer (capacity) as $C=\log_2(1+\rho)$ per frequency, where $\rho$ is the channel's signal-to-noise ratio (SNR)[1]. To meet our information society's ever-expanding quest for more wireless information transfer, the two obvious options of using more bandwidth or improving the SNR both quickly become unfeasible: bandwidth is very limited and expensive, and improving the SNR is a game of rapidly diminishing returns given the logarithmic scaling. Consequently, researchers began to explore the possibility of using multiple channels[2-5] and over the past decades, many intricate approaches to optimally exploit the available wireless channels have been developed. The crux of multichannel systems is always to avoid cross-talk between channels that are not sufficiently independent. The proposed techniques range from elaborate signal coding[6] or beamforming, such as uneven power distribution across transmitters ("water-filling"),[3,7,8] to hardware designs leveraging the orthogonality of different polarization states[9] or combining micro-structured antenna near-fields with time reversal protocols[10]. Ultimately, the propagation medium at hand imposes the upper bound on what any of these techniques can achieve in terms of channel cross-talk. Here, we introduce the idea of physically shaping the propagation medium itself and hence the channels it offers for wireless communication, rather than accepting these as given. The idea is to remove any channel cross-talk by imposing perfect orthogonality of the wireless channels.

To that end, we deploy simple tunable metasurfaces to alter the environment's boundary conditions. We present microwave experiments demonstrating the feasibility of achieving completely independent channels. Our findings show that a propagation medium's channel diversity is not written in stone but can be optimized by tweaking the medium's disorder with simple metasurfaces; without any effort on the transmit or receive sides, perfect channel orthogonality can be imposed, which is beyond what even idealistic randomness can achieve.

To rigorously unravel the role of channel diversity in multiantenna communication, it is instructive to adopt the conventional matrix formalism in which the entry $H_{\alpha\beta}$ of the channel matrix **H** describes the gain between transmitter $\alpha$ and receiver $\beta$. Consider for concreteness a static $n \times n$ system without channel knowledge at the transmitter. Then the generalized version of Shannon's law reads

$$C(\mathbf{H}, \rho) = \log_2[\det(\mathbf{I}+(\rho/n)\mathbf{H}\mathbf{H}^\dagger)] = \Sigma_i \log_2[1+(\rho/n)\sigma_i^2], \tag{1}$$

where **I** is the identity matrix and $\sigma_i$ are the singular values of **H**. It is easy to see that a perfectly orthogonal channel matrix yields the enticing linear scaling of the capacity with $n$, since all of its singular values are equal. On the other hand, if all channels are identical, Eq. 1 collapses to Shannon's original single-channel equation, albeit with improved SNR. Realistic scenarios are between these two extreme cases. In general, the more disordered the propagation medium is, the more distinguishable the channels are. Yet, even true randomness only yields approximately orthogonal channels as $n$ becomes very large.

Quantifying the channel diversity requires a figure of merit that is more elaborate than the rank of **H**, which is not only prone to noise but above all an integer quantity that is unfitting for any optimization. Equation 1 suggests that the singular value distribution contains the necessary information: the flatter it is, the closer is **H** to being orthogonal. A convenient metric for rank optimization is hence the effective rank introduced in ref.[11]:

$$R_{\text{eff}}(\mathbf{H}) = \exp[-\Sigma_i \sigma_i' \ln(\sigma_i')], \tag{2}$$

where $\sigma_i' = \sigma_i / (\Sigma_i \sigma_i)$ are the normalized singular values of **H**. [See Supplementary Discussion in which we show that the distribution of normalized singular values that yields the maximal effective rank also yields the maximal capacity as defined by Eq. 1.] Applied, for instance, to an idealistic, perfectly random 5×5 Rayleigh fading system, we find an average effective rank of 4.1 rather than 5, meaning there are effectively only 4 independent channels.

Having identified a suitable optimization functional, we now move on to the experiment. Common indoor environments constitute disordered cavities of low quality factor for wireless communication signals due to their irregular geometry from the wave's perspective. The wave is reflected off the walls and other obstacles multiple times and a stationary, speckle-like wave field is established[5]. For a first series of experiments, we work in a chaotic metallic cavity as depicted in Fig. 1a [see Methods] because (i) this constitutes a stable, well-controlled system, and (ii) it is easy to perform many realizations with different chaotic geometries while preserving the global physical parameters (volume, quality factor…). It is important to note, however, that the true disorder in this experiment is somewhat contrived relative to realistic environments, such that the channel diversity prior to optimization is already high. The chaotic cavity emulates Rayleigh fading which is considered to provide as much channel diversity as naturally attainable, being *completely* random – unlike most realistic environments. We cover 6% of the cavity walls with a metasurface consisting of 65 electronically reconfigurable elements[12]; each element can mimic Neumann or Dirichlet boundary conditions[13,14] and thereby shape the wave field inside the cavity.

The standard experiment we perform consists in measuring **H** between two simple antenna arrays inside the cavity, and iteratively identifying a metasurface configuration that maximizes $R_{\mathrm{eff}}(\mathbf{H})$ and thereby the channel diversity. [See Methods.] We repeat this experiment for different orientations of the mode-stirrer and different positions of the antennas. In Fig. 1b we show for 2×2, 4×4 and 6×6 systems the evolution of the channel matrix' effective rank over the course of the iterative optimization, both for a sample single realization and the average over 30 realizations. For reference, the values of $R_{\mathrm{eff}}$ expected under Rayleigh fading, and for a perfectly orthogonal channel matrix are indicated. Indeed, we see that at the outset the channel diversity corresponds to Rayleigh fading conditions. For $n=2$, the optimization quickly saturates at the optimum of $R_{\mathrm{eff}}=2$. For $n=4$, the optimum of $R_{\mathrm{eff}}=4$ is also reached, after a few more iterations. For $n=6$, a substantial improvement of the channel diversity is achieved that falls just a little short of the optimum.

To get physical insight into the optimization dynamics, we now analyze the channel matrix in more detail. For a sample realization with $n=4$, we plot in Fig. 2 the evolution of the normalized eigenvalues of **H**. Over the course of the optimization, the initially randomly distributed eigenvalues are forced onto the unit circle in the complex plane and end up having the exact same weight. This is a clear signature of the channel matrix having become perfectly orthogonal. Consequently, cross-talk between the 4 channels has been completely eliminated by physically shaping the channels. In Fig. 3a-c we show the normalized eigenvalue distribution before and after optimization for all 30 realizations. Initially, the eigenvalues are approximately uniformly spread across the unit circle. This is predicted by Girko's full-circle law for random matrices as $n$ becomes large[15]. The corresponding singular value spectra in Fig. 3d-f are downward sloping, in excellent agreement with random matrix theory[3,15]. Note that the slope decreases as $n$ becomes larger; this is expected because large random matrices are approximately orthogonal. After optimization, the eigenvalues in Fig. 3a-c are relocated onto the unit circle's perimeter. While this is achieved perfectly for $n=2$, as $n$ gets larger our limited control over the wave field gradually becomes insufficient and the eigenvalues are scattered around the perimeter within a range that increases with $n$. The tendency to equalize the weights is still clearly visible but eventually not fulfilled perfectly anymore. Correspondingly, the singular value spectra in Fig. 3d-f after optimization are perfectly flat up to $n=4$ but only approximately flat for larger $n$.

Traditional wavefront shaping tools explain why the performance eventually slowly degrades, as $n$ becomes larger. With our fixed-size metasurface, we can effectively control a certain number of cavity modes depending on the modal overlap[16]. The more channels we seek to shape, the more our fixed amount of control over the wave field is shared amongst the channels. As $n>4$, the available control becomes less and less sufficient to attain perfect orthogonality. [See Supplementary Discussion.] Using a larger metasurface or improving its design are two straightforward options to counteract this.

The channel cross-talk directly reflects the discussed physical properties of the channel matrix. As shown in Fig. 3g-i, in all cases there was substantial channel cross-talk before optimization. In line with the slope of the singular value spectrum, the initial channel cross-talk is lower, the larger $n$ is. After optimization, the cross-talk has been completely eliminated up to $n=4$, and has been reduced to a finite but negligible level for larger $n$.

At this stage, having thus far focused our discussion on taming the disorder to yield optimal channel diversity, a question that naturally arises is what impact our channel shaping has on the transmitted energy and thereby on the SNR. Since our antennas only probe a tiny portion of the cavity wave field, the transferred energy before and after optimization does not necessarily have to be the same; by creating constructive or destructive interferences, the transmitted energy balance may in principle vary in either direction[16-18]. This must be seen in contrast to communication via multimode waveguides such as optical fibers[19,20]. However, since our optimization criterion was solely based on orthogonality considerations, one would not expect any significant change of transmitted energy. A convenient quantity to evaluate the latter is $E=\langle \sigma_i^2 \rangle_i$ since we can see from Eq. 1 that any change of $E$ can be directly interpreted as a change in SNR. Indeed, we observe that the ratio of final to initial $E$ is very close to unity in our experiments, the average being 0.94. Due to the logarithmic scaling, small changes in $E$ do not significantly impact the achievable capacity. Nonetheless, one could in principle define alternative optimization functionals that seek trade-offs between channel diversity and SNR.

Finally, to provide both an illustrative visualization of the benefits of optimal channel diversity as well as compelling evidence that our scheme works in real-life, we transpose our experiment from the well-controlled chaotic cavity to a standard office room. [See Methods.] We consider the wireless transfer of the RGB image in Fig. 4a via a simple 3×3 multiantenna system without any coding (one monochrome color component per antenna). First, we measure the initial channel matrix. It has an effective rank of 2.2 rather than 3, meaning there are effectively only two independent channels. Note how this real-life scenario has some channel diversity but remains well below the Rayleigh fading benchmark. Then we optimize the channel diversity as before, attaining a final effective rank of 2.9 which corresponds almost to the optimum of 3. Based on the experimentally measured initial and final channel matrices, we then emulate the image transfer with a binary phase shift keying scheme in combination with regulated zero forcing [see Methods]. We thereby assume channel knowledge only on the receiver side which is routinely obtained with pilot signals. Initially, the transmitted image is heavily distorted (Fig. 4b); in particular, orange and yellow are not distinguishable and white surfaces appear pink. After optimizing the channel diversity, the image is transmitted almost flawlessly (Fig. 4c).

We have demonstrated experimentally the ability of imposing perfect orthogonality of wireless communication channels in (usually to some extent disordered) propagation media by physically shaping the latter with simple metasurfaces, achieving optimal channel diversity and minimal cross-talk. Our concept may find further applications in future chip-to-chip communications where, besides optical solutions, radio-frequency wireless interconnects are currently considered to overcome today's bottle-neck: electrical wires[21-23].

1. Shannon, C. E. A Mathematical Theory of Communication. *Bell Syst. Tech. J.* **27**, 379-423 (1948).
2. Foschini, G. J. & Gans, M. J. On Limits of Wireless Communications in a Fading Environment when Using Multiple Antennas. *Wirel. Pers. Commun.* **6**, 311-335 (1998).
3. Telatar, E. Capacity of Multi-antenna Gaussian Channels. *Trans. Emerg. Telecommun. Technol.* **10**, 585-595 (1999).
4. Moustakas, A. L., Baranger, H. U., Balents, L., Sengupta, A. M. & Simon, S. H. Communication Through a Diffusive Medium: Coherence and Capacity. *Science* **287**, 287-290 (2000).
5. Simon, S. H., Moustakas, A. L., Stoytchev, M. & Safar, H. Communication in a Disordered World. *Phys. Today* **54**, 38-43 (2001).
6. Alamouti, S. M. A simple transmit diversity technique for wireless communications. *IEEE J. Sel. Areas Commun.* **16**, 1451-1458 (1998).
7. Miller, D. A. Establishing Optimal Wave Communication Channels Automatically. *J. Lightwave Technol.* **31**, 3987-3994 (2013).
8. Yan, Y. *et al.* High-capacity millimetre-wave communications with orbital angular momentum multiplexing. *Nat. Commun.* **5**, 4876 (2014).
9. Andrews, M. R., Mitra, P. P. & deCarvalho, R. Tripling the capacity of wireless communications using electromagnetic polarization. *Nature* **409**, 316 (2001).
10. Lerosey, G., de Rosny, J., Tourin, A. & Fink, M. Focusing Beyond the Diffraction Limit with Far-Field Time Reversal. *Science* **315**, 1120-1122 (2007).
11. Roy, O. & Vetterli, M. in *15th European Signal Processing Conference.* 606-610 (IEEE).
12. Kaina, N., Dupré, M., Fink, M. & Lerosey, G. Hybridized resonances to design tunable binary phase metasurface unit cells. *Opt. Express* **22**, 18881-18888 (2014).
13. Sievenpiper, D., Zhang, L., Broas, R. F., Alexopolous, N. G. & Yablonovitch, E. High-impedance electromagnetic surfaces with a forbidden frequency band. *IEEE Trans. Microwave Theory Tech.* **47**, 2059-2074 (1999).
14. Sihvola, A. Metamaterials in electromagnetics. *Metamaterials* **1**, 2-11 (2007).
15. Tulino, A. M. & Verdú, S. *Random Matrix Theory and Wireless Communications*. Vol. 1 (2004).
16. Dupré, M., del Hougne, P., Fink, M., Lemoult, F. & Lerosey, G. Wave-Field Shaping in Cavities: Waves Trapped in a Box with Controllable Boundaries. *Phys. Rev. Lett.* **115**, 017701 (2015).
17. del Hougne, P., Rajaei, B., Daudet, L. & Lerosey, G. Intensity-only measurement of partially uncontrollable transmission matrix: demonstration with wave-field shaping in a microwave cavity. *Opt. Express* **24**, 18631-18641 (2016).
18. Kaina, N., Dupré, M., Lerosey, G. & Fink, M. Shaping complex microwave fields in reverberating media with binary tunable metasurfaces. *Sci. Rep.* **4**, 6693 (2014).
19. Ambichl, P. *et al.* Super-and Anti-Principal-Modes in Multimode Waveguides. *Phys. Rev. X* **7**, 041053 (2017).
20. Böhm, J., Brandstötter, A., Ambichl, P., Rotter, S. & Kuhl, U. In situ realization of particlelike scattering states in a microwave cavity. *Phys. Rev. A* **97**, 021801 (2018).
21. Miller, D. A. Device Requirements for Optical Interconnects to Silicon Chips. *Proc. IEEE* **97**, 1166-1185 (2009).

22  Chang, M. F., Roychowdhury, V. P., Zhang, L., Shin, H. & Qian, Y. RF/wireless interconnect for inter-and intra-chip communications. *Proc. IEEE* **89**, 456-466 (2001).
23  Phang, S. *et al*. Near-Field MIMO Communication Links. *IEEE Trans. Circuits Syst. I, Reg. Papers* (2018).
24  Vellekoop, I. M. & Mosk, A. Phase control algorithms for focusing light through turbid media. *Opt. Commun.* **281**, 3071-3080 (2008).
25  Popoff, S., Lerosey, G., Fink, M., Boccara, A. C. & Gigan, S. Image transmission through an opaque material. *Nat. Commun.* **1**, 81 (2010).

**Acknowledgments** P.d.H. thanks Alexandre Aubry for fruitful discussions. P.d.H. acknowledges funding from the French "Ministère de la Défense, Direction Générale de l'Armement".

**Author Contributions** G.L. initiated the project. P.d.H. conceived and performed the experiments, analyzed the data and wrote the manuscript. All authors discussed the project and the results.

**Competing Interests** The authors declare no competing interests.

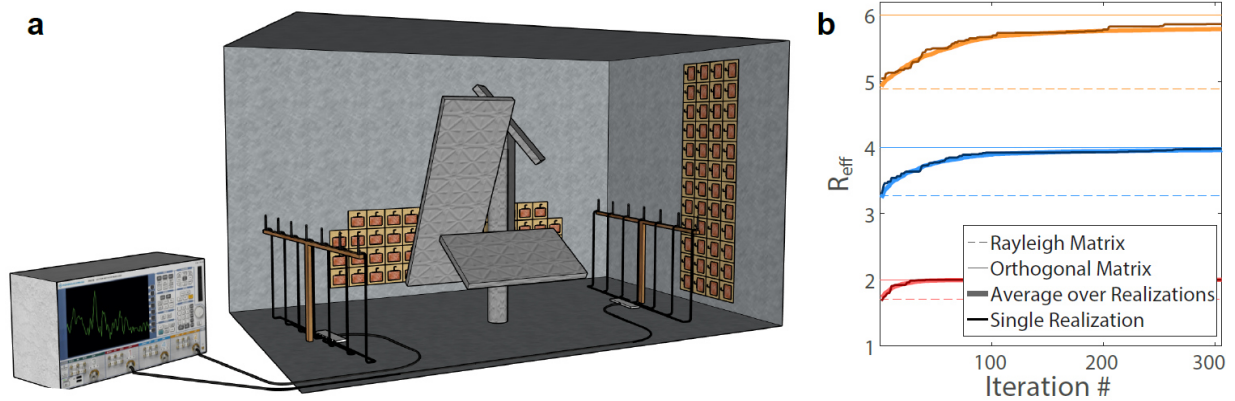

**Figure 1 | Experimental setup and procedure. a,** Disordered, metallic cavity under Rayleigh fading conditions. A phase-binary metasurface reflect-array partially covers the cavity walls; appropriately configured, it physically shapes the channel matrix measured between the two antenna arrays and imposes perfect channel orthogonality. **b,** Iterative optimization of channel diversity. The evolution of $R_{\mathrm{eff}}$ over the course of the optimization is given for a single realization, as well as averaged over 30 realizations, for $n=2,4,6$ (red, blue, yellow). Benchmarks for Rayleigh fading and perfect orthogonality are indicated, see legend.

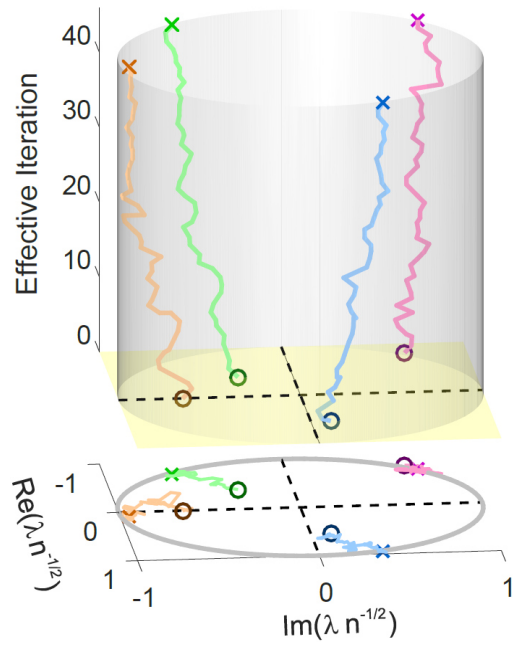

**Figure 2 | Evolution of channel matrix eigenvalues during iterative optimization of channel diversity.** Only "effective iterations" are shown, i.e. the ones that increase $R_{\text{eff}}$. Data corresponds to a single experimental realization of a 4×4 system.

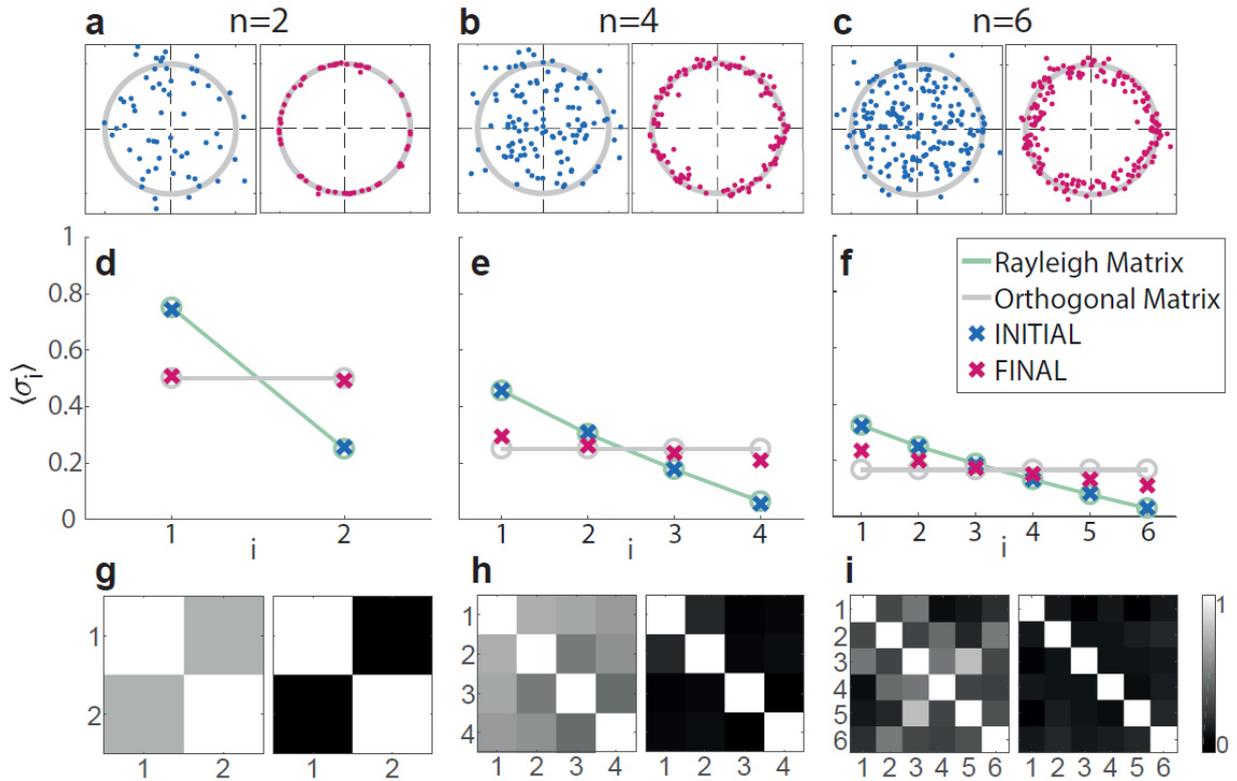

**Figure 3 | Orthogonality of optimized channel matrices. a-c,** Distribution of normalized eigenvalues of the experimentally measured channel matrices, before (left) and after (right) taming the disorder, for $n=2,4,6$. The unit circle is indicated for reference. **d-f,** Normalized singular value spectra of the experimentally measured channel matrices for $n=2,4,6$, averaged over 30 realizations. **g-i,** Sample cross-talk matrices, before (left) and after (right) optimization, for the same individual realizations as in Fig. 1b.

**a** Original Image  **b** Initial (R$_{eff}$ = 2.2)  **c** Final (R$_{eff}$ = 2.9)

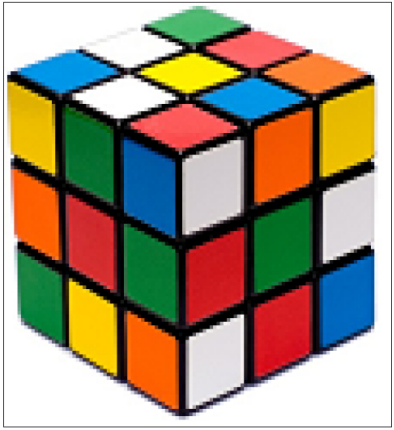 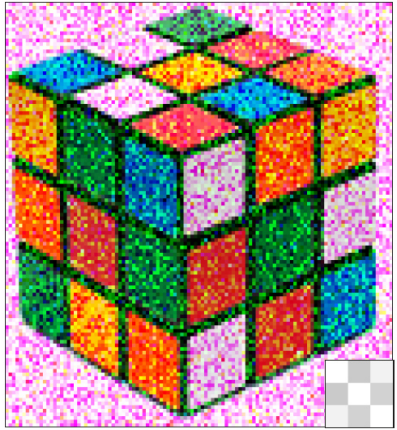 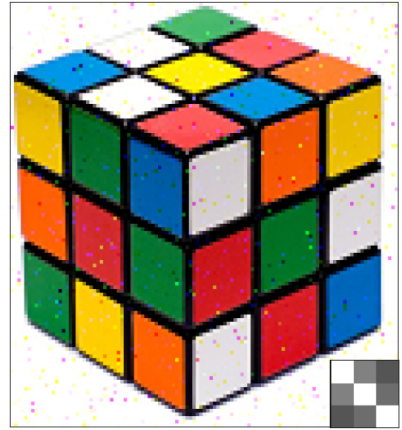

**Figure 4 | Emulated wireless transfer of a full-color image based on experimentally measured channel matrices in an office room.** See SM for detailed procedure. **a,** Original image. **b,** Recovered image with the initial channel matrix. **c,** Recovered image with the optimized channel matrix. Insets show cross-talk matrices corresponding to experimentally measured initial and final channel matrix (same colors as in Fig. 3g-i).

# METHODS

**Experimental Setup and Procedure**

This section provides the technical details of our proof-of-principle experiment in the chaotic metallic cavity shown schematically in Fig. 1a of the main text. The results presented in Fig. 1b, Fig. 2 and Fig. 3 are obtained with this setup.

The cavity dimensions are 1.45 m × 1 m × 0.75 m; the cavity thus has a volume of 1.1 m$^3$. The cavity surface is metallic (aluminum). The ensemble of metallic elements in the cavity (original cavity's walls plus irregular objects in the cavity such as the mode-stirrer) yields a clearly disordered geometry.

The metasurface consists of 102 phase-binary elements; for each element the phase shift of the reflected wave can be tuned electronically via a bias voltage from an Arduino microcontroller to be either 0 or $\pi$. The working principle of the metasurface reflect-array is outlined in detail in ref.[12]. In summary, each element consists of two resonators that hybridize. The resonance of one of the two resonators can be altered with the bias voltage of a PIN diode. Thus the overall resonance of the element can be tuned. The element is designed such that it is either on or off resonance at the working frequency of $f_0$ = 2.47 GHz which is a standard Wi-Fi operation frequency. This yields the phase-binary control over the wave field. The reflect-array covers about 6% of the cavity surface. Note that out of the 102 reflect-array elements 37 were broken such that effectively only 65 were used throughout this work.

We estimated the quality factor $Q$ of the cavity at the working frequency as 604. We evaluated $Q$ as $\pi f_0/\mu$, where $\mu$ is the exponential decay constant extracted from the inverse Fourier transform of the experimentally measured transmitted field, averaged over different random reflect-array configurations and mode-stirrer orientations.

Each array of antennas consisted of a line of 8 identical commercial Wi-Fi monopole antennas separated by 10 cm (a little more than half the wavelength at the working frequency). All 8 antennas were in the exact same orientation (i.e. no polarization diversity). Each antenna array was connected to a radio-frequency switch which in turn was connected with a vector network analyzer.

The optimization procedure was a standard iterative sequential algorithm[24]. We used the effective rank of the channel matrix (see Eq. 2 in main text) as cost function. For each iteration, we changed the configuration of one of the reflect-array elements, measured the new channel matrix and evaluated the new effective rank. If it was higher than previously, we updated the optimal reflect-array configuration accordingly. This way, we tested element after element, looping three times over each element. Examples of experimental optimization curves are given in Fig. 1b in the main text. Retesting each element multiple times is necessary because the reverberation inside the cavity creates long-range correlations between the optimal configurations of the reflect-array elements. Indeed, the optimization curves for $n=4$ and $n=6$ of the example experiments displayed in Fig. 1b in the main text do not saturate after having tested each reflect-array element once (iteration #65) but significantly later.

Finally, we performed 30 realizations of the experiment for each considered square channel matrix size $n \in \{2,3,4,5,6,7\}$. A realization of disorder may be defined as repeating the experiment with the same global parameters (cavity volume, quality factor …) but with a different (disordered) cavity geometry. We ran the experiment for ten different mode-stirrer orientations and three different random choices of $n$ antennas amongst the 8 available ones.

**Numerical Evaluation of Rayleigh Fading Channel Model**
The model of Rayleigh fading channels assumes an i.i.d. complex, zero mean, unit variance distribution of the entries of the channel matrix. This is satisfied by picking both the real and the imaginary part of each entry of the channel matrix from a normally distributed random number generator. The energy normalization is irrelevant for the evaluation of the effective rank. For each value of $n$, we simulated 500 random Rayleigh fading channel matrices in MatLab and computed the corresponding quantities (effective rank, singular value spectrum).

**Evaluation of Channel Cross-Talk**
The entry $O_{ij}$ of the overlap matrix $\mathbf{O}$ quantifies how the vector $A_i$ of channel gains from every transmitter to receiver $i$ overlaps with the vector $A_j$ of channel gains from every transmitter to receiver $j$. We normalize these vectors by the root-mean-square of the absolute values of their entries, yielding $A_i'$ and $A_j'$. Then, $O_{ij}=|A_i'^{\dagger}A_j'|$ and we define the average channel overlap as the average of all off-diagonal entries of $\mathbf{O}$. A perfectly orthogonal matrix has $O_{ij} = \delta_{ij}$, where $\delta_{ij}$ is the Kronecker delta.

**Office Room Implementation**
To transpose our experiment to real-life, we replaced the well-controlled and perfectly stable metallic cavity with a standard office room. The room's dimensions are 2.5 m × 4.3 m × 3.6 m; its volume is thus 38.7 m$^3$ and our reflect-array covers only 0.5% of the room's surface. We measured a quality factor of 215 in the room. The room is furnished with multiple tables, chairs and cupboards. A photographic image of the room is provided in Extended Data Figure 1. Throughout the experiments, there was no human motion in the room. The experimental procedure was exactly the same as before, the only difference being that the metallic disordered cavity was replaced by a furnished office room. A sample optimization curve obtained experimentally in the office room for a 3×3 system is displayed in Extended Data Figure 2, as well as the singular value spectrum of the initial and optimized channel matrix. This data corresponds to the realization on which Fig. 4 of the main text is based.

**Emulating a Wireless Image Transmission with Experimentally Measured Channels**
Here we detail the procedure used to produce Fig. 4 of the main text. Based on experimentally measured wireless communication channels in an office room (see above), we emulated the transfer of an RGB color image across a 3×3 Multiple-Input Multiple-Output system ($n$=3). Thanks to the linearity of wave propagation, all these calculations can be performed numerically based on the experimentally measured channels. To produce Fig. 4, we chose $\rho$ = 6.8 dB. The objective is to visually illustrate the benefits of shaping wireless channels. Hence, we focus on the Physics of the problem and use a simple modulation/demodulation scheme rather than a more elaborate one. We do explicitly *not* claim that the scheme is optimal, but it serves very well our purpose of visualizing the impact of improved channel diversity on a wireless image transmission.

We consider a simple binary phase shift keying (BPSK) scheme. Moreover, we do not assume channel knowledge at the transmitter. At the receiver, the channel is assumed to be known, which is usually achieved through the use of pilot signals which are followed by the actual data in practical schemes. Furthermore, we assume that the channel is static throughout the transmission of the image.

An RGB image is a three-dimensional matrix, the third dimension being of length three which corresponds to the three color components red (R), green (G) and blue (B). Each matrix entry is an integer between 0 and 255 which corresponds to 8 bits. We transform the information for each of the three colors into a one-dimensional binary vector. To that end, we reshape for each color the corresponding two-dimensional matrix into a vector and then replace each entry by eight consecutive entries corresponding to the entry's binary representation.

This yields three binary one-dimensional data streams. We assign one of the three transmitting antennas to each data stream and simultaneously emit one entry of the data streams at a time. In the BPSK scheme, we assign '1' to a phase of 0, and '0' to a phase of $\pi$. We obtain the three received signals by simply multiplying the transmitted signals with the experimentally measured channel matrix and adding the desired level of white Gaussian noise. There is thus no effort whatsoever neither on the design of the antenna arrays, nor on the transmitted signals, nor their power.

To recover the data from the received signals, we opt for a simple linear data processing scheme on the receiver side. Given the measurement Y=**H**X+n and the channel matrix **H**, the challenge is to recover X. Note that except for the very special case of an identity matrix as channel matrix, there is *always* the need for some data processing – irrespective of the channel diversity. A simple but effective technique is that known as 'minimum mean square error', or also referred to as 'regulated zero forcing'. It consists in estimating X as $([\mathbf{H}^\dagger \mathbf{H}+\beta \mathbf{I}]^{-1} \mathbf{H}^\dagger)$ Y, where † denotes the conjugate transpose operation. $\beta$ is the standard deviation of the experimental noise. The benefit of this technique is that it works irrespective of the SNR. In fact, for very low or very high SNR it becomes equivalent to simply using the inverse channel matrix, i.e. $\mathbf{H}^{-1}$, or using phase conjugation, i.e. $\mathbf{H}^\dagger$, respectively. We do not delve into the details here but refer the interested reader to ref.[25] which provides an insightful in-depth discussion and detailed analysis of the technique. The key point about choosing the value of $\beta$ is that it directly depends on the SNR; in a realistic experiment $\beta$ can thus be chosen once and for all. Here, we thus simply test a wide range of values and pick the best one.

Once we have estimated **X** this way, we recast the data streams into the original binary form. If an entry's phase is between $-\pi/2$ and $\pi/2$, we identify a '1', else a '0'. This yields once again three one-dimensional binary data streams like X, and we apply the inverse procedure to reconvert them into a two-dimensional 8-bit RGB image. The quality of the reconstruction can be evaluated, for instance as the correlation coefficient between the true and the recovered image. We used this as cost function to identify the optimal value of $\beta$.

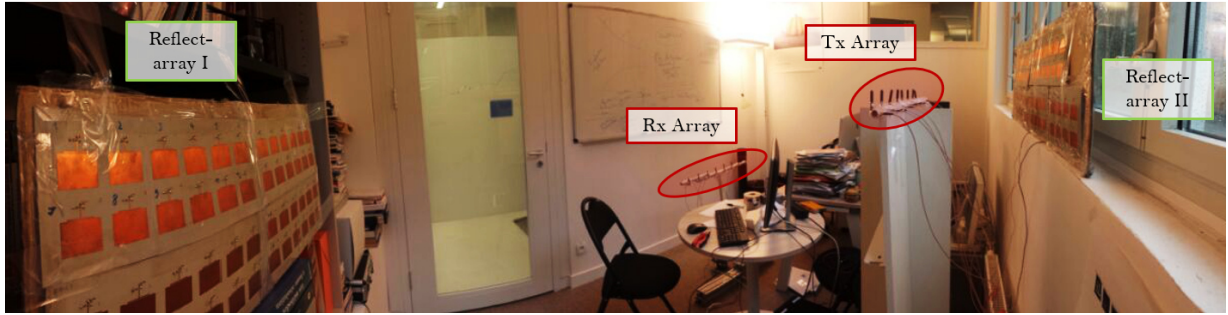

**Extended Data Figure 1 | Photographic image of office room setup.** Two antenna arrays are placed in a heavily furnished office of irregular geometry. The two parts of the metasurface reflect-array are also identified in the image.

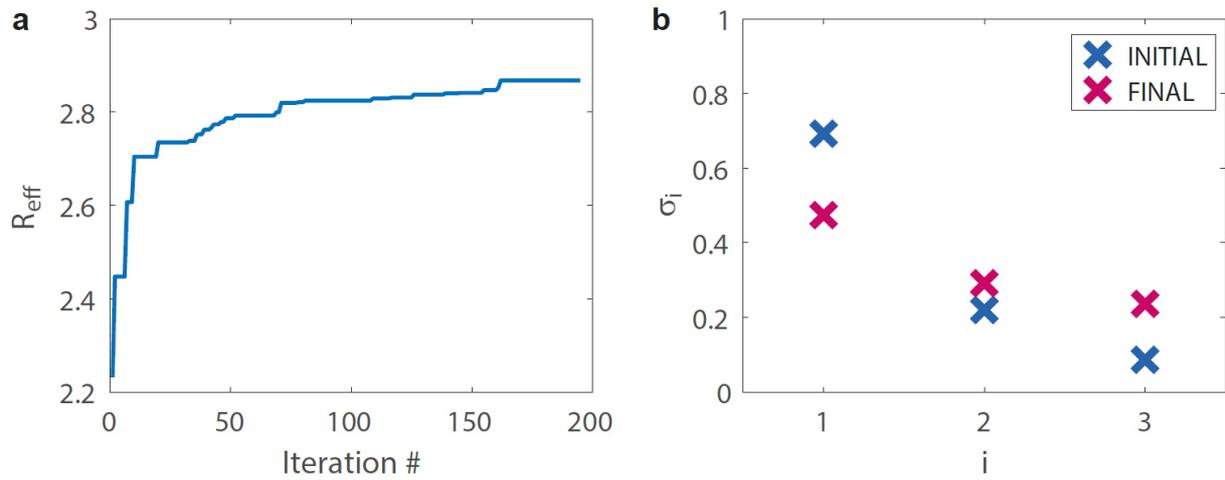

**Extended Data Figure 2 | Further details on the realization of the experiment in the office room.** Figure 4 of the main text is based on this experiment. The office room is shown in Extended Data Figure 1. **a,** Optimization dynamics as function of the iteration. **b,** Singular value spectrum of the channel matrix before and after optimization in the office room.

**SUPPLEMENTARY INFORMATION:**

**Optimal Communication Channels in a Disordered World**

**with Tamed Randomness**

Philipp del Hougne[1]*, Mathias Fink[1], Geoffroy Lerosey[2]*

[1]Institut Langevin, CNRS UMR 7587, ESPCI Paris, PSL Research University, 1 rue Jussieu, 75005 Paris, France.

[2]Greenerwave, ESPCI Paris Incubator PC'up, 6 rue Jean Calvin, 75005 Paris, France.

*Correspondence to philipp.delhougne@gmail.com or geoffroy.lerosey@greenerwave.com.

**Supplementary Discussion**

*Performance under Real-Life Conditions*

Compared with the metallic cavity, the office room has a substantially larger volume as well as a significantly lower quality factor. The impact of these two factors on the ability of the reflect-array to shape the wireless channels can be understood with traditional wavefront shaping tools. As mentioned in the main text and detailed in refs.[1-3], the control of the metasurface reflect-array over the wave field may be modeled as effectively controlling some $p$ modes out of the $N$ cavity modes that overlap at the working frequency due to their finite line-widths. Based on Weyl's Law, $N$ may be estimated as

$$N = \frac{8\pi f_0^3}{c^3} \times \frac{V}{Q},$$

where $c$ is the speed of light and $V$ is the cavity volume. The term $\frac{V}{Q}$ increased by a factor of about 141 between the relatively small, high-$Q$ metallic cavity and the very large, low-$Q$ office room. Consequently, $N$ is a lot larger in the office room than in the metallic cavity, while $p$ remains the same since we still use the same reflect-array. Therefore, our control over the wave field is naturally a lot smaller in the office room than in the metallic cavity – given our experimental setup. There are several obvious ways to increase the reflect-array's control over the wave field. First, one may simply use a larger reflect-array with more elements. Indeed, in our office room experiment we only covered a tiny percentage of the wall surface with our reflect-array. Second, one may improve the reflect-array design. Currently, it only has a phase-binary control over a single polarization and moreover suffers from attenuation issues; there is ample room for refining the technology.

The optimization in the office room clearly does not reach perfect orthogonality ($R_\mathrm{eff} = 2.86 \neq n = 3$) which can also be seen by the final singular value spectrum in Extended Data Figure 2b not being perfectly flat. As discussed above, a notable deterioration in the optimization performance is expected in the office room as compared with the metallic cavity. Nonetheless, the channel diversity is substantially enhanced even in the office room, as visualized in Fig. 4 of the main text.

Finally, we point out that the entire optimization procedure could be performed in real time with optimized electronics (e.g. using field-programmable-gate-arrays (FPGAs) instead of the Arduino microcontroller) such that the reflect-array's PIN diodes could be switched at MHz rates. Hardware solutions in that spirit have been reported, for instance, in refs.[4,5], proving that this is technologically possible with the current state-of-the-art.

*Algebraic Proof that Capacity is Maximized by Singular Values of Equal Weight*

In this section, we provide a mathematical proof for what is intuitively obvious and stated in the main text: the achievable information transfer capacity as defined in Eq. 1 of the main text is maximal if all singular values of the channel matrix have the same weight.

First, we recall Eq. 1 in terms of the square singular values of the channel matrix:

$$C = \sum_{i=1}^{n} \log_2\left(1 + \frac{\rho}{n} s_i\right),$$

where $s_i = \sigma_i^2$ is the square of the $i$th singular value $\sigma_i$ of **H**.

We wish to maximize $C(s_1, s_2, \cdots, s_n)$ subject to the constraint $\sum_{i=1}^{n} s_i = P$, where $P$ is some positive, real-valued constant. Now, we apply the method of Lagrange multipliers. The Lagrangian for the problem at hand is

$$\mathcal{L}(s_1, s_2, \cdots, s_n, \zeta) = C(s_1, s_2, \cdots, s_n) + \zeta\, g(s_1, s_2, \cdots, s_n),$$

with $g(s_1, s_2, \cdots, s_n) = \left(\sum_{i=1}^{n} s_i\right) - P$. The gradients can thus be evaluated as

$$\frac{\partial \mathcal{L}}{\partial s_i} = \frac{\partial}{\partial s_i}\left(\log_2\left(1 + \frac{\rho}{n} s_i\right) + \zeta s_i\right) = \frac{\rho}{s_i \rho \ln(2) + n \ln(2)} + \zeta$$

and

$$\frac{\partial \mathcal{L}}{\partial \zeta} = \left(\sum_{i=1}^{n} s_i\right) - P$$

yielding

$$\frac{\rho}{s_i \rho \ln(2) + n \ln(2)} + \zeta = 0$$

and

$$\left(\sum_{i=1}^{n} s_i\right) - P = 0,$$

the latter being the original constraint.

By considering the constraints for $s_i$ and $s_j$, for $1 \leq i \leq n$, $1 \leq j \leq n$ and $i \neq j$,

$$\frac{\rho}{s_i \rho \ln(2) + n \ln(2)} + \zeta = \frac{\rho}{s_j \rho \ln(2) + n \ln(2)} + \zeta$$

which simplifies to

$$s_i = s_j.$$

Hence, to maximize $C(s_1, s_2, \cdots, s_n)$ subject to the constraint $\sum_{i=1}^{n} s_i = P$, all $s_i$ have to be identical, quod erat demonstrandum.

*Algebraic Proof that Effective Rank is Maximized by Singular Values of Equal Weight*

In this section, we provide a mathematical proof that the effective rank of the channel matrix is maximized if the singular values of the channel matrix are equal. In conjunction with the previous section, this result rigorously justifies our choice of the effective rank as optimization functional in order to maximize the capacity of the channel matrix.

We proceed as in the previous section by using the method of Lagrange multipliers. Here, we wish to maximize

$$R_{\text{eff}}(\sigma_1', \sigma_2', \cdots, \sigma_n') = \exp\left(-\sum_{i=1}^{n} \sigma_i' \ln(\sigma_i')\right) = \prod_{i=1}^{n} e^{-\sigma_i' \ln(\sigma_i')},$$

where $\sigma_i' = \sigma_i / (\sum_{i=1}^{n} \sigma_i)$ are the normalized singular values of the channel matrix (see Eq. 2 in the main text), subject to the constraint $\sum_{i=1}^{n} \sigma_i' = 1$. The Lagrangian for the problem at hand is

$$\mathcal{L}(\sigma_1', \sigma_2', \cdots, \sigma_n', \zeta) = R_{\text{eff}}(\sigma_1', \sigma_2', \cdots, \sigma_n') + \zeta\, g(\sigma_1', \sigma_2', \cdots, \sigma_n'),$$

with $g(\sigma_1', \sigma_2', \cdots, \sigma_n') = (\sum_{i=1}^{n} \sigma_i') - 1$. The gradients can thus be evaluated as

$$\frac{\partial \mathcal{L}}{\partial \sigma_i'} = -\sigma_i'^{-\sigma_i'}(\ln(\sigma_i') + 1)\left(\prod_{j=1, j\neq i}^{n} e^{-\sigma_j' \ln(\sigma_j')}\right) + \zeta$$

and

$$\frac{\partial \mathcal{L}}{\partial \zeta} = \left(\sum_{i=1}^{n} \sigma_i'\right) - 1.$$

This yields

$$0 = -\sigma_i'^{-\sigma_i'}(\ln(\sigma_i') + 1)\left(\prod_{j=1, j\neq i}^{n} e^{-\sigma_j' \ln(\sigma_j')}\right) + \zeta$$

and

$$0 = \left(\sum_{i=1}^{n} \sigma_i'\right) - 1.$$

The latter is once again the original constraint. From the former, it follows for $1 \leq i \leq n$, $1 \leq k \leq n$ and $i \neq k$, that

$$-\sigma_i'^{-\sigma_i'}(\ln(\sigma_i') + 1)\left(\prod_{j=1, j\neq i}^{n} e^{-\sigma_j' \ln(\sigma_j')}\right) + \zeta = -\sigma_k'^{-\sigma_k'}(\ln(\sigma_k') + 1)\left(\prod_{j=1, j\neq k}^{n} e^{-\sigma_j' \ln(\sigma_j')}\right) + \zeta$$

which simplifies to

$$\sigma_i'^{-\sigma_i'}(\ln(\sigma_i') + 1)e^{-\sigma_k' \ln(\sigma_k')} = \sigma_k'^{-\sigma_k'}(\ln(\sigma_k') + 1)e^{-\sigma_i' \ln(\sigma_i')}$$

and then to

$$\frac{\ln(\sigma_i') + 1}{\sigma_i'^{\sigma_i'}} e^{\sigma_i' \ln(\sigma_i')} = \frac{\ln(\sigma_k') + 1}{\sigma_k'^{\sigma_k'}} e^{\sigma_k' \ln(\sigma_k')}$$

which ultimately may be recast as

$$\ln(\sigma_i') = \ln(\sigma_k').$$

This implies $\sigma_i' = \sigma_k'$. Hence, to maximize the effective rank subject to the constraint $\sum_{i=1}^{n} \sigma_i' = 1$, all normalized singular values $\sigma_i'$ have to be identical, quod erat demonstrandum.

1 Dupré, M., del Hougne, P., Fink, M., Lemoult, F. & Lerosey, G. Wave-Field Shaping in Cavities: Waves Trapped in a Box with Controllable Boundaries. *Phys. Rev. Lett.* **115**, 017701 (2015).
2 Kuhl, U., Stöckmann, H. & Weaver, R. Classical wave experiments on chaotic scattering. *J. Phys. A* **38**, 10433 (2005).
3 Hemmady, S., Zheng, X., Antonsen Jr, T. M., Ott, E. & Anlage, S. M. Universal statistics of the scattering coefficient of chaotic microwave cavities. *Phys. Rev. E* **71**, 056215 (2005).
4 Yang, H. *et al.* A programmable metasurface with dynamic polarization, scattering and focusing control. *Sci. Rep.* **6**, 35692 (2016).
5 Gollub, J. *et al.* Large Metasurface Aperture for Millimeter Wave Computational Imaging at the Human-Scale. *Sci. Rep.* **7**, 42650 (2017).